\documentclass[%
 aps,
 prl,
 reprint,
 superscriptaddress,
]{revtex4-1}

\usepackage{graphicx}
\usepackage{dcolumn}
\usepackage{bm}

\def\apj{Astrophys. J.}
\def\apjl{Astrophys. J.\ letts.}

\def\jgr{J. \ Geophys. Res.}

\def\pp{Phys. Plasmas}

\def\prl{Phys. Rev. Lett.}

\let \aap = \aa

\begin{document}

\title{Energetics of kinetic reconnection in a three-dimensional null points cluster}

\author{V. Olshevsky}
\email[]{vyacheslav.olshevsky@wis.kuleuven.be}
\affiliation{Center for mathematical Plasma Astrophysics (CmPA)
Department of Mathematics, KU Leuven, Celestijnenlaan 200B, 
bus 2400 B-3001 Leuven, Belgium.}
\affiliation{Main Astronomical Observatory of NAS, 
Akademika Zabolotnoho 27, 03680, Kyiv, Ukraine}
\author{G. Lapenta}
\affiliation{Center for mathematical Plasma Astrophysics (CmPA)
Department of Mathematics, KU Leuven, Celestijnenlaan 200B, 
bus 2400 B-3001 Leuven, Belgium.}
\author{S. Markidis}
\affiliation{High Performance Computing and Visualization (HPCViz), 
KTH Royal Institute of Technology, SE-100 44, Stockholm, Sweden}


\begin{abstract}
We performed three-dimensional Particle-in-Cell simulations of magnetic 
reconnection with multiple magnetic null points. 
Magnetic field energy conversion into kinetic energy was about five times 
higher than in traditional Harris sheet configuration. 
More than $85\%$ of initial magnetic field energy was transferred to 
particle energy during $25$ reversed ion cyclofrequencies. 
Magnetic reconnection in the cluster of null points evolved in three phases. 
During the first phase, ion beams were excited, 
that then gave part of their energy back to magnetic field in the second phase.
In the third phase, magnetic reconnection occurs in many small patches around 
the current channels formed along the stripes of low magnetic field.
Magnetic reconnection in null points presents essentially three-dimensional 
features, with no two dimensional symmetries or current sheets. 
\end{abstract}


\maketitle

Null-point magnetic reconnection is thought to be the main source of energy 
release in solar flares, Earth magnetosphere and other astrophysical plasmas. 
Simulations of 3D null point reconnection in magnetohydrodynamic (MHD) approach 
were reported in \cite{Galsgaard:Nordlund:1997JGR,Galsgaard:Pontin:2011a,
Galsgaard:Pontin:2011b}; null-point reconnection regimes were reviewed and 
classified by \cite{Priest:Pontin:2009PhPl}; in the papers 
\cite{Lau:Finn:1990,Pontin:etal:2005GApFD} 3D null point reconnection was 
studied in kinematic approach; \cite{Dalla:Browning:2005,Guo:etal:2010,Stanier:etal:2012} 
investigated particle acceleration in the vicinity of 3D null-points.
In many astrophysical applications, such as planet magnetospheres or solar 
flares, fluid models are not sufficient to describe magnetic reconnection 
\cite{Ji:Daughton:2011,Baalrud:2011}. 
To our knowledge, only \cite{Baumann:Nordlund:2012ApJ} reported 
Particle-in-Cell (PIC) simulations of an isolated null point in solar corona. 
To understand the basic kinetic and 3D effects of null point reconnection,
we have performed PIC simulation of a cluster of null points in the simplest
possible cofiguration that excludes the complications due to boundary condition 
or external sources. 
We have found that energy release in such system is much
more efficient than in traditional Harris sheet configurations. 
Three-dimensional effects dominate in the dynamics throughout the simulation, 
and no current sheets or other 2D structures are formed during reconnection.
We believe that both laminar and turbulent reconnection regimes were realised
in our experiment, characterized by different energy dissipation and particle 
acceleration 
\cite{Biskamp:1986,Lazarian:Vishniac:1999,Lapenta:2008PhRvL}.

%
%
\begin{figure*}
\includegraphics[width=0.99\textwidth]{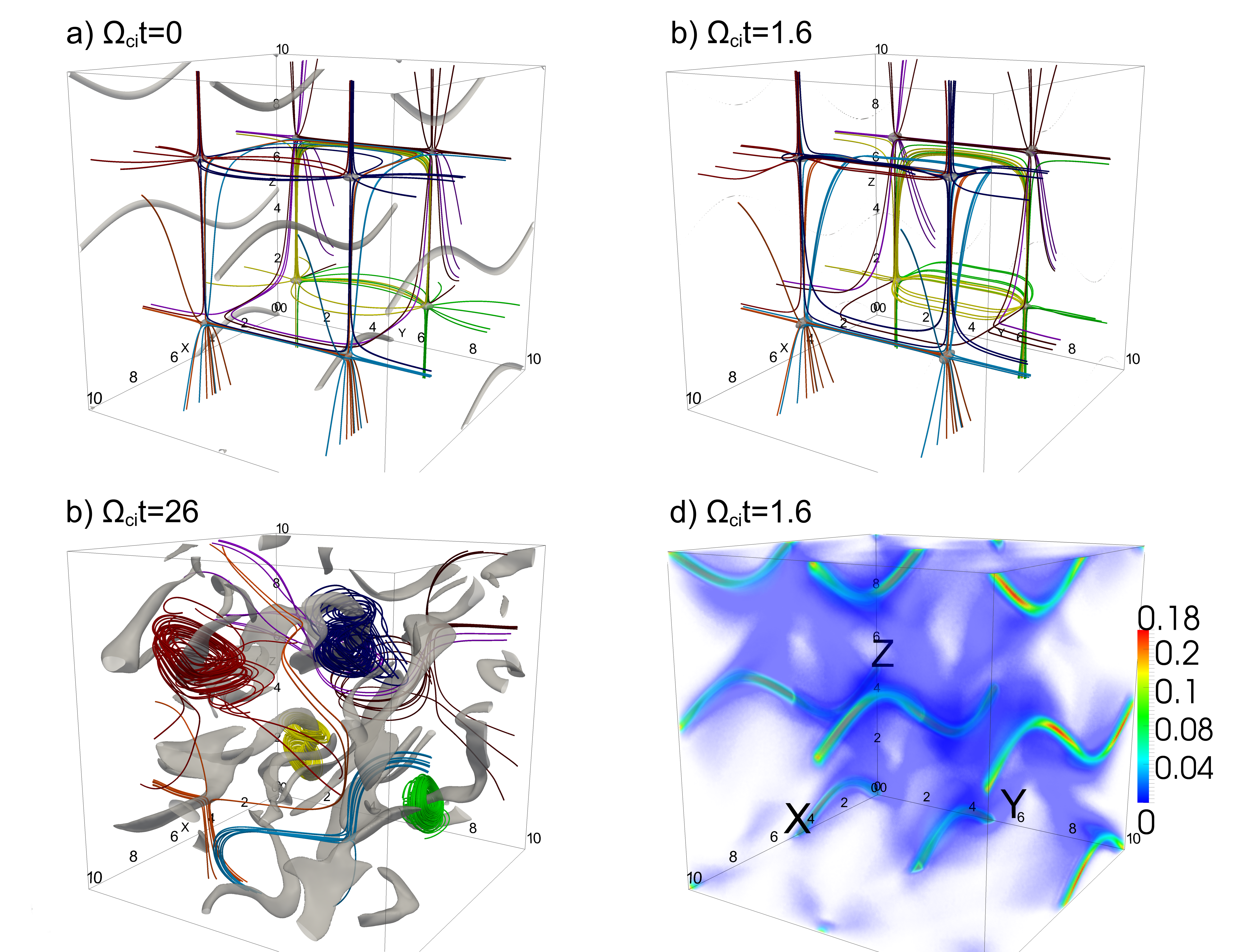}
\caption{
{\bf a, b, c}: Snapshots of magnetic field at different times 
(labelled in the upper left). Field lines are plotted in different colors; 
each color corresponds to the lines passing by one of $8$ initial null points. 
Isocontours of low magnetic field $B=0.1 B_0$ are plotted in grey.
{\bf d}: Snapshot of current $J$ made at $\Omega_{ci}t=1.6$.
\label{fig:field}
}
\end{figure*}
Making the very first step towards kinetic models of 3D null point reconnection, 
we designed our simulations to be as simple and straightforward, as possible.
In particular, we impose boundaries in our simulation domain periodic in all 
directions. 
In a fully periodic box, it can be proved that the minimum number of points where 
magnetic field vanishes, is $8$.
We have chosen a non-equilibrium initial state with uniformly 
spaced $8$ null points:
\begin{eqnarray*}
  B_x = -B_0\cos(\frac{2\pi x}{L_x})\sin(\frac{2\pi y}{L_y}), \qquad\qquad \\
  B_y = B_0\cos(\frac{2\pi y}{L_y})\left( \sin(\frac{2\pi x}{L_x}) - 
        2 \sin(\frac{2\pi z}{L_z}) \right), \\
  B_z = 2 B_0\sin(\frac{2\pi y}{L_y})\cos(\frac{2\pi z}{L_z}), \qquad\qquad \\
\end{eqnarray*}
where $B_0$ is the magnetic field amplitude; $L_x$, $L_y$, and $L_z$ are the 
sizes of the simulation domain in the corresponding directions.
It is easy to show that this configuration satisfies the condition 
$\mathbf{\nabla}\cdot\mathbf{B}=0$. Besides $8$ null points, 
the condition $B=0$ also holds along $9$ ``null lines'' lied up in the 
planes $Y=\pi n$, where $n$ is an integer (Fig.~\ref{fig:field}).

The collisionless plasma simulations were carried out using iPIC3D, 
fully kinetic electromagnetic PIC code with implicit time stepping 
\cite{markidis:etal:2010}.
Simulation domain represents a cubic box with dimensions 
$L_x\times L_y\times L_z$ = $10\,d_i\times10\,d_i\times 10\,d_i$
with $256^3$ cells and $64$ particles of each specie per cell. 
Our plasma consisted of electrons and ions with mass ratio 
$m_i/m_e=25$ and temperature ratio $T_i/T_e=5$; 
the particles were initialized with Maxwellian velocity distribution 
with electron thermal velocity $\upsilon_{the}/c=0.0346$. 
The initial particle density was uniform with $n_0=1$; 
the initial magnetic field amplitude was $B_0=0.03$.
It is important to note, that with such parameters the electron spatial scales 
are well resolved in our simulation: the electron skin depth 
$d_e= 0.2 \, d_i = 5 \Delta x$,  where $\Delta x$ is the grid spacing.
The time step was set relative to ion plasma frequency: 
$\Delta t = 0.125/\omega_{pi}$.
For consistency, in the following all times are normalized to ion cyclofrequency 
computed with the initial field amplitude 
$\Omega_{ci}=eB_0/m_ic=0.052\, \omega_{pi}$.

Our simulation domain initially includes null points of both A and B types 
\cite{Cowley:1973}, as shown in Figure~\ref{fig:field}.
All the null points are interconnected with each other by separators, 
formed by the intersections of separatrix surfaces and spines. 
Many of these separators are the artifacts of the highly 
symmetric configuration, and must not necessarily exist in nature 
\cite{Greene:1988}; they are destroyed very quickly after beginning of the 
simulation.

At $\Omega_{ci}t=1.6$ (panel b in Fig.~\ref{fig:field}) no aforementioned 
``null line'' is visible, and the connectivity of many field lines has 
changed.
Initially, gas and magnetic pressures are not balanced, and the magnetic fields 
surrounding the ``null lines'' are compressed by plasma. 
Particles move towards these regions and establish 
the current channels that dominate the evolution of the system, 
as shown in panel d of Figure~\ref{fig:field}. 
The energy released by magnetic field in this phase of the simulation is 
transferred mainly to ions accelerated in the current channels.
The radius of the cross-section of such channel is of the order of $d_i$.
Current channels become unstable later in time, they are continuously distorted, 
and multiple stripes of small $B$ intensity are formed along them.
As we will demonstrate below, the reconnection sites are associated with 
these regions.
There is no initial current through the null points in our configuration, 
and the null points do not move during the simulation.
They are destroyed in pairs (see article supplemented material).

At the end of the simulations all $8$ null points are destroyed, and 
the magnetic field organization is not reminiscent of the initial configuration.
The magnetic field lines spiral around current channels established in the 
regions of low $B$.
Note, that we have terminated the simulation at time $\Omega_{ci}t=26$ 
when most, but not all the magnetic field lines are reconnected.
All the features forming during the simulation are essentially 
three-dimensional, and current sheets or other 2D structures are not created.

%
\begin{figure}
\includegraphics[width=\columnwidth]{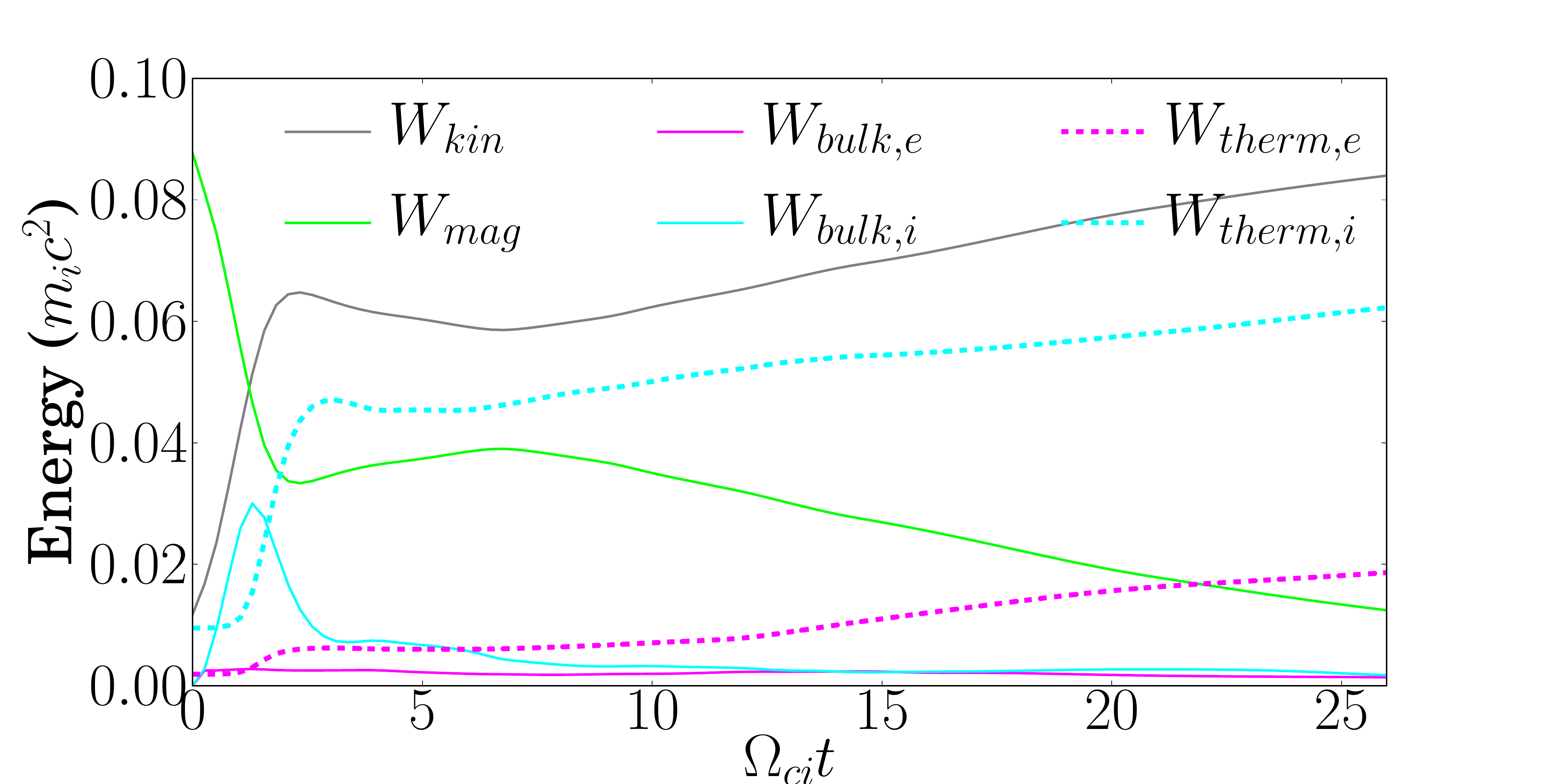}
\caption{
Evolution of different components of energy (total kinetic energy of particles, 
magnetic field energy, bulk and thermal energy of both species). 
\label{fig:energy}
}
\end{figure}
\begin{figure*}
\includegraphics[width=0.99\textwidth]{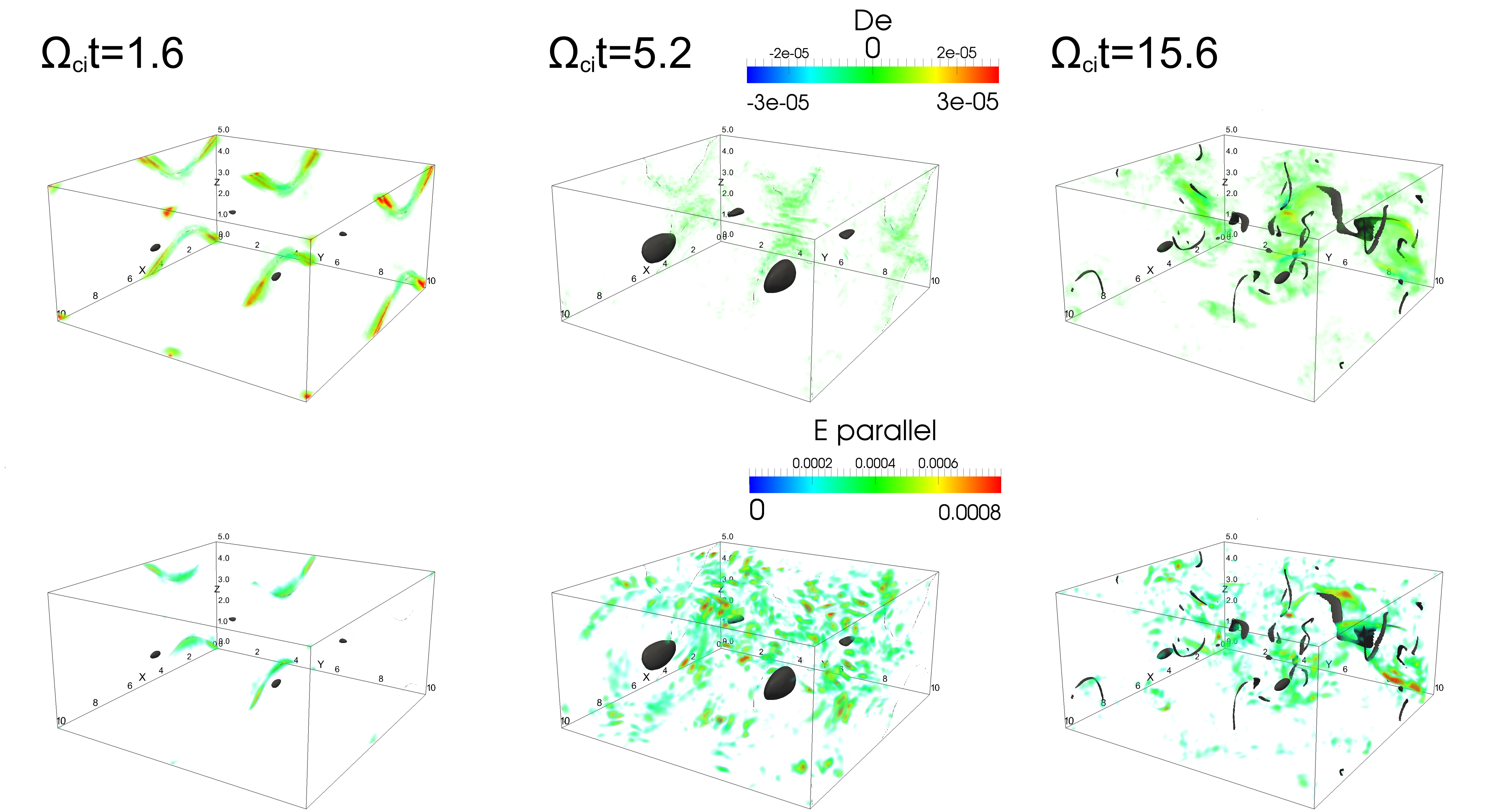}
\caption{
Snapshots of the electron-frame dissipation measure $D_e$ (upper panels) 
and parallel electric field $E_\|$ (lower panels) in the lower half of the 
computational domain spanning from $Z=0$ to $Z=5\,d_i$ at different times 
(labelled in the upper left of each column). 
Isocontours of low magnetic field $B=0.1B_0$ are plotted in grey.
\label{fig:diss}
}
\end{figure*}
Figure~\ref{fig:energy} shows the evolution of the magnetic field energy 
$W_{mag}$, total kinetic energy of the particles $W_{kin}$, bulk and thermal 
kinetic energy for species ($W_{bulk,e}$, $W_{bulk,i}$, $W_{therm,e}$, 
and $W_{therm,i}$, correspondingly).
The electric field energy is two orders of magnitude smaller than other 
components of energy, and is not shown in the plot. 
Initially, the magnetic energy contributes $85\%$ to the total energy of the 
system; particles contribute only $12\%$. 
At the end of the simulation, particle energy constitutes $87\%$ of the total 
energy: $86\%$ of magnetic energy is released within $26\Omega_{ci}^{-1}$!
This corresponds to decay rate 
$\lambda=\left(\Delta W_{mag}/\Delta t\right)/W_{mag} \approx 0.033\Omega_{ci}$,
which is almost $5$ times higher than $\lambda\approx 0.0075\Omega_{ci}$ 
observed in ``classical'' two-dimensional Harris current sheet 
reconnection setup \cite{Markidis:etal:2012b,Vapirev:etal:2013}.

In the beginning of the simulation, when the current channels are established, 
most of the released magnetic energy goes to the energy of ion bulk motions 
$W_{bulk,i}$. 
Very soon, though, these motions are slowed down, and in the later phase of 
reconnection the temperature of species is increased rather than fluid motions.
At the end of the simulations temperature ratio reduces to $T_i/T_e\approx 3$,
and fluid motions constitute only a few percent to the total energy in the 
system.

As in \cite{Wan:etal:2012,Zenitani:etal:2011}, we compute the electron-frame 
dissipation measure
\begin{equation}
  D_e = \mathbf{J}\cdot 
    \left( \mathbf{E} - \left[\mathbf{J}_e\times\mathbf{B}\right]/\rho_e \right)
    + \rho_c\mathbf{u}_e\cdot\mathbf{E},
\label{eq:diss}
\end{equation}
where $\mathbf{J}$ and $\mathbf{J}_e$ are the total and electron current 
densities, correspondingly, $\rho_e$ and $\mathbf{u}_e$ are electron density 
and speed, and $\rho_c=e(n_i-n_e)$ is charge density.
The dissipation $D_e$ characterizes the work done by fields on particles 
$\mathbf{J}\cdot\mathbf{E}$ with corrections for convective motions 
$\mathbf{J}\cdot[\mathbf{J}_e\times\mathbf{B}]/\rho_e$, and for charge 
separation $\rho_c\mathbf{u}_e\cdot\mathbf{E}$.
The latter quantity is by orders of magnitude smaller than two other components 
of $D_e$ in our simulations.
The first two components balance each other so that $D_e$ is very small.

For additional illustration, we have selected the ``lower'' half of our 
computational domain that spans from $Z=0$ to $Z=5 \,d_i$, 
and contains $4$ null points and sections of $6$ ``null lines''.
Figure~\ref{fig:diss} shows snapshots of $D_e$ and parallel electric field 
$E_{\|}=\mathbf{E}\cdot \mathbf{B}/B$ in this slab.

The magnetic reconnection dynamics may be divided into three distinct phases 
according to the energy evolution. 
The first phase of rapid acceleration of particles and very high dissipation 
$D_e$ lasts till $\Omega_{ci}t\approx 2$. 
During this time powerful currents are established in the low $B$ value channels.
The speeds of particle beams in these channels are up to $3.3V_A$ for electrons and 
$1.6V_A$ for ions, where $V_A$ is Alfv\'en speed. 
Most of the electromagnetic energy at this stage goes to accelerate ion fluid 
motions; electron acceleration is much smaller.
The dissipation regions and the regions of enhanced $E_\|$ surround the current 
channels (left column in Fig.~\ref{fig:diss}).

During the second phase of magnetic reconnection, 
$2.5 \lesssim \Omega_{ci}t \lesssim 7$, 
work of the electromagnetic field on particles is negative, 
i.e., particles give some of their energy back to the field.
The main contributor of this ``reverse'' energy exchange is the deceleration 
of ion currents.
Dissipation areas surround the current channels, while the regions of large 
values of $E_{\|}$ are scattered randomly throughout the domain, 
so there is no correlation between these locations 
(middle column in Fig.~\ref{fig:diss}).
The value of $T_i/T_e\approx 3$ establishes in this phase, 
and does not substantially vary later.

In the last phase of magnetic reconnection, small-scale reconnection 
events are chaotically distributed in the regions of low $B$ 
(right column in Fig.~\ref{fig:diss}). 
The locations of enhanced $D_e$ and $E_\|$ show well correspondence to 
each other.
This phase is characterized by a steady growth of particle energy and a 
decrease of the magnetic energy (Fig.~\ref{fig:energy}).
The released magnetic energy accelerates particle random motions and 
increase their effective temperature.
The steady (turbulent) reconnection process continues until all the null 
points are destroyed and all field lines reconnect (not shown in this paper).


To summarize, we have performed PIC simulations of magnetic reconnection 
in a 3D domain that contains $8$ null points.
The simulation was terminated at $\Omega_{ci}t=26$, when all the initial 
null points vanish, and the majority of magnetic field lines spiral
around the regions of low magnetic field and powerful electric currents.
During this period, $86\%$ of initial magnetic field energy was transferred 
to particle kinetic energy, and the decay rate was almost $5$ times higher 
than in two-dimensional Harris sheet reconnection.
All the magnetic and current features that formed during the simulations 
were essentially three-dimensional, no current sheets or other 2D structure 
were identified.

Three phases of magnetic reconnection could be distinguished: 
(i) a rapid establishment of powerful current channels, when most of energy 
is gained by ion beams;
(ii) a negative dissipation phase, when ion beams slow down and return some 
energy back to electromagnetic field; 
(iii) a ``steady'' reconnection phase, characterized by multiple small-scale 
reconnection events localized around the current channels. 
The first two phases lasted for $\Omega_{ci}t\approx2$, 
and $\Omega_{ci}t\approx5$, correspondingly, 
while the third one was still developing when the simulation was terminated.
Finally, most of the magnetic energy was converted to the thermal energy 
of the particles, so that ion/electron temperature ratio changed from $5$ 
in the beginning to $3$ in the end of simulations.

Interestingly, during the second phase the topology of magnetic field 
was changing, but the dissipation regions did not coincide with the 
regions of enhanced parallel electric field. 
During the ``steady'' reconnection, however, these locations showed 
good correspondence to each other, surrounding the regions where the 
reconnection took place.

\begin{acknowledgments}
Authors are grateful to Dr. Clare Parnell for the useful comments on the 
structure of the magnetic configuration.
This research has received funding from the European Commission's FP7 
Program with the grant agreement SWIFF (project 2633430, swiff.eu).
The simulations were conducted on the computational resources provided 
by the PRACE Tier-0 project 2011050747 (Curie supercomputer)
\end{acknowledgments}

\bibliography{3dnull}

\end{document}